\documentclass[runningheads]{llncs}
\usepackage[T1]{fontenc}
\usepackage{amsfonts}
\usepackage{graphicx}
\newcommand{\parder}[2]{\frac{\partial #1}{\partial #2}}

\newcommand{\tparder}[2]{\partial #1/\partial #2}

\begin{document}
\title{A relation between $k$-symplectic and $k$-contact Hamiltonian system}
%
%
\author{S. Vilariño\inst{1}\orcidID{0000-0003-0404-1427
} 
}
\authorrunning{S. Vilariño}
%
\institute{Department of Applied Mathematics and Institute of Mathematics and Applications (IUMA), University of Zaragoza.  Torres Quevedo Building, Campus Rio Ebro, 50018, Zaragoza, Spain.\\
\email{silviavf@unizar.es}\\
}
\maketitle              
\begin{abstract}
Systems of partial differential equations which appear in classical field theories can be studied geometrically using different geometrical structures, for example, $k$-symplectic geometry, $k$-cosymplectic geometry, multisymplectic geometry, etc. 
In recent years, there has been a notable increase in the study of $k$-contact Hamiltonian systems. These are based on the description of the dynamics of field theories using the so-called $k$-contact manifolds. Such structures are generalizations of contact structures and $k$-symplectic structures. 
The relation between $k$-symplectic manifolds and $k$-contact manifolds was established in \cite{LRS24}. In light of the above relation, this work seeks to explore the relationship between $k$-symplectic Hamiltonian systems and $k$-contact Hamiltonian systems.

\keywords{Hamiltonian systems \and $k$-contact manifolds \and $k$-symplectic manifolds.}
\end{abstract}

\section{Introduction}

There are several geometrical settings for studying systems of partial differential equations that appear in field theories. Of particular interest in recent years is the study of $k$-contact structures \cite{LRS24}, which can be considered a simultaneous generalization of the contact structures of mechanics and the $k$-symplectic structures from the study of classical field theories. 
The study of these structures has revealed a relationship between the $k$-symplectic and $k$-contact manifolds. Specifically,  an exact $k$-symplectic structure, $(P, \mathbf{\theta})$, induces a canonical $k$-contact structure on $M=P\times\mathbb{R}^k$, called the contactification of a $k$-symplectic manifold.

The aim of this work is to analyze the relationship between $k$-symplectic Hamiltonian $k$-vector fields and $k$-contact Hamiltonian $k$-vector fields when we consider a Hamiltonian $k$-symplectic system and the induced Hamiltonian $k$-contact system. As a result of this study, it is concluded that it is possible to recover the solutions of $k$-vector fields of the $k$-symplectic Hamiltonian system from the solutions of the induced $k$-contact Hamiltonian system. Finally we present the vibrating string problem as an example of the results of this work.

To ensure consistency and clarity, we establish some general assumptions and notation that will be used throughout this work, unless otherwise specified. It is hereby assumed that all structures are smooth and well-defined over the reals. Manifolds are Hausdorff, connected, paracompact, and finite-dimensional. It is assumed that closed differential forms have constant-rank. Summation over crossed repeated indices is comprehensible, though it can be explicitly detailed at times for clarity. All our considerations are local to stress our main ideas and to avoid technical problems concerning the global manifold structure of quotient spaces and similar issues. Natural numbers are assumed to be greater than zero.

\section{A briefly review of $k$-symplectic and $k$-contact manifolds}

This section reviews the fundamental notions and results on $k$-symplectic and $k$-contact geometry that will be used later for the description of the respective Hamiltonian systems (see \cite{LRS24,GGMRR_20,GGMRR_21} for more details). 

In the following, $\mathfrak{X}(P)$ and $\Omega^k(P)$ stand for the $\mathcal{C}^\infty(P)$-modules of vector fields and differential $k$-forms on a manifold $P$. The canonical basis of $\mathbb{R}^k$, denoted as $\{e_1,\ldots,e_k\}$, gives rise to the dual basis $\{e^1,\ldots, e^k\}$ in $\mathbb{R}^{k*}$. 

Given an $n$-dimensional manifold $M$ we consider the Whitney sum over $M$ of $k$ copies of the tangent bundle, namely $\bigoplus^k TM :=  TM\oplus_M\ldots\oplus_M  TM$ ($k$ times)
with the natural projections $pr_M\colon\bigoplus^k TM\to M$ and $pr_M^\alpha\colon\bigoplus^k TM\to TM$ with $\alpha=1,\ldots, k$, where $pr^\alpha_M$ denotes the projection from $\bigoplus^k TM$ onto its $\alpha$-th component and $pr_M$ is the projection from $\bigoplus^k TM$ onto the base manifold $M$.

Each $\mathbf{\theta}\in \Omega^\ell(P,\mathbb{R}^k)$ can be represented as $\mathbf{\theta}=\theta^\alpha\otimes e_\alpha$ for some uniquely defined differential forms $\theta^1,\ldots,\theta^k \in \Omega^\ell(P)$. 

\subsection{$k$--vector fields and integral sections}

Let us review the theory of $k$-vector fields, which plays a crucial role in the geometric analysis of systems of partial differential equations \cite{LSV15}. 

\begin{definition}
A {\it $k$-vector field} on $M$ is a section $\mathbf{X}\colon M\to\bigoplus^k TM$ of the vector bundle $pr_M\colon\bigoplus^k TM\rightarrow M$. The space of $k$-vector fields on $M$ is denoted by $\mathfrak{X}^k(M)$.
\end{definition}

It is important to observe that each $k$-vector field $\mathbf{X}\in\mathfrak{X}^k(M)$ is equivalent to a family of $k$ vector fields $X_1,\ldots,X_k\in\mathfrak{X}(M)$, defined by $X_\alpha = pr_M^\alpha\circ\mathbf{X}$ for $\alpha=1,\ldots,k$. This fact justifies denoting $\mathbf{X}:= (X_1,\ldots, X_k)$. 

\begin{definition}\label{dfn:first-prolongation-k-tangent-bundle}
   Given a map $\phi\colon U\subset\mathbb{R}^k\to M$, its {\it first prolongation} is the map $\phi^{(1)}\colon U\subset\mathbb{R}^k\to\bigoplus^k TM$ 
    defined as follows
    $$ \phi^{(1)}(t) = \left( \phi(t);  T _t\phi\left( \parder{}{t^1}\bigg\vert_t \right),\ldots, T _t\phi\left( \parder{}{t^k}\bigg\vert_t \right) \right)
    \,,$$ where $t=(t^1,\ldots,t^k)\in\mathbb{R}^k\,.
    $
\end{definition}


\begin{definition}
    Let $\mathbf{X} = (X_1,\ldots,X_k)\in\mathfrak{X}^k(M)$ be a $k$-vector field. An {\it integral section} of $\mathbf{X}$ is a map $\phi\colon U\subset\mathbb{R}^k\to M$ such that $\phi^{(1)} = \mathbf{X}\circ\phi\,$, namely $ T\phi \left(\parder{}{t^\alpha}\right) = X_\alpha\circ\phi$ for $\alpha=1,\ldots,k$. A $k$-vector field $\mathbf{X}\in\mathfrak{X}^k(M)$ is {\it integrable} if $[X_\alpha,X_\beta]=0$ for $\alpha,\beta=1,\ldots,k$.
\end{definition}

Let $\mathbf{X} = (X_1,\ldots, X_k)$ be a $k$-vector field on $M$ with local expression $ X_\alpha = X_\alpha^i\parder{}{x^i}\,$ for $\alpha=1,\ldots ,k$.
Then, $\phi\colon U\subset\mathbb{R}^k\to M$ is an integral section of $\mathbf{X}$ if, and only if, its coordinates satisfy the following system of PDEs
\begin{equation}
\label{Eq::IntegralSections}
	\parder{\phi^i}{t^\alpha} = X_\alpha^i\circ\phi\,,\qquad i=1,\ldots,n,\qquad \alpha=1,\ldots,k\,.
\end{equation}

\subsection{{$k$}--Symplectic geometry}

This section briefly surveys the theory of $k$-symplectic manifolds (see \cite{LSV15}).

\begin{definition}
A {\it $k$-symplectic form} on $P$ is a closed nondegenerate\footnote{$\omega=\omega^\alpha\otimes e_\alpha$ is nondegenerate if and only if $\cap_{\alpha=1}^k\ker\omega^\alpha=\{0\}.$} $\mathbb{R}^k$-valued two-form $\mathbf{\omega}$. The pair $(P,\mathbf{\omega})$ is called a {\it $k$-symplectic manifold}. If $\mathbf{\omega}$ is exact, namely $\mathbf{\omega} = \mathrm{d}\mathbf{\theta}$ for some $\mathbf{\theta} \in\Omega^1(P,\mathbb{R}^k)$, then the pair $(P,\mathbf{\theta})$ is namely an {\it exact $k$-symplectic manifold}.
\end{definition}

\begin{definition}
 A \textit{polarised $k$-symplectic manifold} is a triple $(P,\mathbf{\omega}, \mathcal{V})$ such that $(P,\mathbf{\omega})$ is a $k$-symplectic manifold of dimension $n+nk$ and there exists an integrable distribution $\mathcal{V}$ of rank $nk$ with ${\mathbf{\omega}}\vert_{\mathcal{V}\times \mathcal{V}} =0$.
\end{definition}

\begin{theorem}\label{Th:kSymTh}($k$-symplectic Darboux theorem)
Let $(P,\mathbf{\omega}, \cal V)$ be a polarised $k$-symplectic manifold. There exists a local coordinate system around each $x\in P$, given by $\{x^i,y^\alpha_i\}$ with $i=1,\ldots, n,$ and $\alpha=1,\ldots,k$, such that 
\[
{\mathbf{\omega}}=  \mathrm{d} x^i\wedge \mathrm{d} y^\alpha_i\otimes e_\alpha\,,\qquad  \mathcal{V} =\left\langle \frac{\partial}{\partial y^\alpha_i}\right\rangle_{{i=1,\ldots,n\,,\,\alpha=1,\ldots,k}}\,.
\]
\end{theorem}
The coordinates $\{x^i,y^\alpha_i\}$ in the $k$-symplectic Darboux theorem are called {\it $k$-symplectic Darboux coordinates} or an {\it adapted coordinate system}. 

\begin{example}[Canonical model for polarised exact $k$-symplectic manifolds]\label{ex:canonical-model-k-symplectic}
    Let $Q$ be a manifold of dimension $n$. A coordinate system $\{q^i\}$ in $Q$ induces a natural coordinate system $\{q^i,p_i^\alpha\}$ in $T^*_kQ:=\bigoplus^kT^* Q$. Consider the canonical form $\theta\in\Omega^1(T^* Q)$ in  $T^*Q$ satisfying $\omega = -d\theta\in\Omega^2(T^* Q)$. Hence,  $T^*_k Q$ has the canonical forms taking values in $\mathbb{R}^k$ given by     
    $\mathbf{\theta}_{T^*_kQ} = (pr^\alpha_Q)^\ast\theta\otimes e_\alpha$ and $ \mathbf{\omega}_{T^*_kQ} = -d\mathbf{\theta}_k\,, $   
    which, in adapted coordinates $\{q^i,p^\alpha_i\}$ for $T^*_kQ$, read
    \[
    \mathbf{\theta}_{T^*_kQ}= p_i^\alpha\mathrm{d} q^i\otimes e_\alpha \ ,\qquad \mathbf{\omega}_{T^*_kQ} = \mathrm{d} q^i\wedge\mathrm{d} p_i^\alpha\otimes e_\alpha \,.
    \]
    Taking all this into account, $\big(T^*_kQ,\mathbf{\omega}_{T^*_kQ},{\cal V}_{T^*_kQ}\big)$, with ${\cal V}_{T^*_kQ} = \ker Tpr_Q$, is a polarised exact $k$-symplectic manifold.  The coordinates $\{q^i,p_i^\alpha\}$ in $T^*_kQ$ are {\it $k$-symplectic Darboux coordinates}. 
\end{example}


\subsection{$k$--Symplectic Hamiltonian systems}

This subsection recalls the formalism for $k$-symplectic Hamiltonian systems and $k$-symplectic Hamilton--De Donder--Weyl geometric equations \cite{LSV15}.


\begin{definition}\label{dfn:k-sympl-hamiltonian-system}
    A {\it $k$-symplectic Hamiltonian system} is a family $(P,\mathbf{\omega},h)$, where $(P,\mathbf{\omega})$ is a $k$-symplectic manifold and $h\in\mathcal{C}^\infty(P)$ is called a {\it $k$-symplectic Hamiltonian function}. The \textit{$k$-symplectic Hamilton--De Donder--Weyl equation}  is the equation 
    \begin{equation}\label{eq::k-sym-HDW-FE}
       \displaystyle\sum_{\alpha=1}^k\iota_{X_\alpha}\omega^\alpha=\mathrm{d} h\,.
    \end{equation}
    for a $k$-vector field $\mathbf{X}=(X_1,\ldots, X_k)\in\mathfrak{X}^k(P)$.
    

    A $k$-vector field $\mathbf{X}$ that satisfies equation (\ref{eq::k-sym-HDW-FE}) for some $h\in\mathcal{C}^\infty(P)$ is called a \textit{$k$-symplectic Hamiltonian $k$-vector field} of the $k$-symplectic Hamiltonian system $(P,\mathbf{\omega}, h).$ 
    It is important to note that the solution of the equation (\ref{eq::k-sym-HDW-FE}) is not unique, (for more details see \cite{LSV15}).
\end{definition}

In Darboux coordinates, equation (\ref{eq::k-sym-HDW-FE}) implies that a
$k$-symplectic Hamiltonian $k$-vector field $\mathbf{X} = (X_1,\ldots, X_k)\in\mathfrak{X}^k(P)$ can be written
\[
X_\alpha=\displaystyle\parder{h}{p_i^\alpha}\displaystyle\frac{\partial}{\partial q^i}+(X_\alpha)^\beta_i\displaystyle\frac{\partial}{\partial p^\beta_i}\,,\;  \alpha=1,\ldots,k\,.\]
 where the functions $(X_\alpha)^\beta_i$ satisfying  $(X_\alpha)^\alpha_i = - \displaystyle\parder{h}{q^i}$.


 A $k$-symplectic Hamiltonian $k$-vector field is not necessarily integrable, but to obtain the solutions of the Hamilton--De Donder--Weyl equations, the existence of integral sections is relevant.

\begin{proposition}
    Let $(P,\mathbf{\omega}, h)$ be a polarized $k$-symplectic Hamiltonian system and consider an integrable $k$-symplectic Hamiltonian $k$-vector field $\mathbf{X}$. If $\psi\colon \mathbb{R}^k\to M$ is an integral section of $\mathbf{X}$, then $\psi$ is a solution of \emph{the Hamilton--De Donder--Weyl field equations} given by the following system of partial differential equations \cite{LSV15}
    \begin{equation}\label{eq:local_HDW_field_eq}
         \parder{h}{p_i^\alpha}\bigg\vert_{\psi(x)}=\parder{\psi^i}{x^\alpha}\bigg\vert_{x}\,,\qquad
        \parder{h}{q^i}\bigg\vert_{\psi(x)}=-\parder{\psi^\alpha_i}{x^\alpha}\bigg\vert_{x} \,.
 \end{equation}

\end{proposition}
 
\subsection{$k$--Contact geometry}

This section briefly surveys the theory of $k$-contact manifolds \cite{GGMRR_20}.

Consider a differential one-form $\eta\in\Omega^1(M)$. Then, $\eta$ spans a smooth co-distribution $\mathcal{C} = \langle\eta\rangle = \{\langle\eta_x\rangle\mid x\in M\}\subset T^* M$. Then, $\mathcal{C}$ has rank one at every point where $\eta$ does not vanish. The annihilator of $\mathcal{C}$ is the distribution $\ker \eta\subset TM$. The distribution $\mathcal{C}^\circ$ has corank one at every point where $\eta$ does not vanish and zero otherwise.

\begin{definition}\label{dfn:k-contact-manifold}
    A \textit{$k$-contact form} on an open $U\subset M$ is a differential one-form on $U$ taking values in $\mathbb{R}^k$, let us say $\mathbf{\eta} = \sum_\alpha\eta^\alpha\otimes e_\alpha \in\Omega^1(U,\mathbb{R}^k)$, such that $1.\,$ $0\neq\ker \mathbf{\eta}\subset TU$ is a regular distribution of corank $k$, $2.\,$$\ker d\mathbf{\eta}\subset TU$ is a regular distribution of rank $k$,
        and $3.\,$$\ker \mathbf{\eta}\cap\ker d\mathbf{\eta}  = 0$.
 A pair $(M,\mathbf{\eta})$ is called a \textit{$k$-contact manifold}. If, in addition, $\dim M = n+nk+k$ for some $n,k\in\mathbb{N}$ and $M$ is endowed with an integrable distribution $\mathcal{V}\subset \ker \mathbf{\eta}$ with $dim\mathcal{V} = nk$, we say that $(M,\mathbf{\eta},\mathcal{V})$ is a \textit{polarised    $k$-contact manifold}. We call $\mathcal{V}$ a \textit{polarisation} of $(M,\mathbf{\eta})$.
\end{definition}



\begin{theorem}\label{thm:k-contact-Reeb}
    Let $(M,\mathbf{\eta})$ be a $k$-contact manifold. There exists a unique family of vector fields $R_1, \ldots, R_k\in\mathfrak{X}(M)$, called the \textit{Reeb vector fields} of $(M, \mathbf{\eta})$, such that
    \begin{equation}\label{eq:k-contact-Reeb}
            \iota_{R_\alpha}\eta^\beta = \delta_\alpha^\beta\,,\qquad
            \iota_{R_\alpha}d\mathbf{\eta} = 0\,,
    \end{equation}
    for $\alpha,\beta = 1,\ldots,k$. Moreover, $[R_\alpha,R_\beta] = 0$ with $\alpha,\beta = 1,\ldots,k$, while $\ker d\mathbf{\eta}=\langle R_1,\ldots, R_k \rangle$, and $R_1\wedge \ldots \wedge R_k$ is non-vanishing.
\end{theorem}


\begin{example}\label{ex:canonical-k-contact-structure}(Canonical example of a polarised $k$-contact manifold)
    The manifold $M = (\bigoplus^kT^* Q)\times\mathbb{R}^k$ has a natural $k$-contact form
    $
        \mathbf{\eta}_{Q}= \sum_{\alpha=1}^k(\mathrm{d} z^\alpha - \theta^\alpha)\otimes e_\alpha\,,
    $
    where $\{z^1,\ldots,z^k\}$ are the pull-back to $M$ of standard linear coordinates in $\mathbb{R}^k$ and each $\theta^\alpha$ is the pull-back of the Liouville one-form $\theta$ of the cotangent manifold $T^* Q$ with respect to the projection $M\to T^* Q$ onto the $\alpha$-th component of $\bigoplus^kT^* Q$. Note also that $M$ admits a natural projection onto $Q\times \mathbb{R}^k$ and a related vertical distribution $\mathcal{V}$ of rank $k\cdot\dim Q$ contained in $\ker \mathbf{\eta}_{Q}$. Hence, $\left((\bigoplus^kT^* Q)\times\mathbb{R}^k,\mathbf{\eta}_Q,\mathcal{V}\right)$ is a polarised    $k$-contact manifold.


\end{example}

\begin{example}[Contactification of an exact $k$-symplectic manifold]\label{ex:contactification-k-symplectic-manifold}
Considering  similar ideas to those described in the previous example one can define a $k$-contact manifold from an exact $k$-symplectic manifold.

    Let $(P,\mathbf{\omega} = \mathrm{d}\mathbf{\theta} )$ be an exact $k$-symplectic manifold and consider the product manifold $M = P\times\mathbb{R}^k$. Let $\{z^1,\ldots,z^k\}$ be the pull-back to $M$ of some Cartesian coordinates in $\mathbb{R}^k$ and denote  by $\theta_M^\alpha$ the pull-back of $\theta^\alpha$ to the product manifold $M$. Consider the $\mathbb{R}^k$-valued one-form ${\mathbf{\eta}} = \sum_{\alpha=1}^k(\mathrm{d} z^\alpha + \theta_M^\alpha)\otimes e_\alpha \in\Omega^1(M,\mathbb{R}^k)$.  Then, $(M,\mathbf{\eta})$ is a    $k$-contact manifold, because $\ker \mathbf{\eta}\neq 0$ has corank $k$, while $d{\mathbf{\eta}} = d\mathbf{\theta}_M$ and $\ker d{\mathbf{\eta}} = \langle\tparder{}{z^1},\dots,\tparder{}{z^k}\rangle$ has rank $k$ since $\mathbf{\omega}$ is non-degenerate. It follows that $\mathbf{\eta}$ is a globally defined $k$-contact form.

\end{example}

\begin{theorem}[$k$-contact Darboux Theorem \cite{GGMRR_20}]
\label{Th::PolkCon}
    Consider a polarised $k$-contact manifold $(M, \mathbf{\eta}, \mathcal{V})$. Then, around every point of $M$, there exist local coordinates $\{q^i,p_i^\alpha,z^\alpha\}$, with $1\leq\alpha\leq k$ and $1\leq i\leq n$, such that
    \[
    \mathbf{\eta} = \left(\mathrm{d} z^\alpha - p_i^\alpha\mathrm{d} q^i\right)\otimes e_\alpha\, , \quad \ker d\mathbf{\eta} = \left\langle \parder{}{z^\alpha}\right\rangle\,,\quad \mathcal{V} = \left\langle\parder{}{p_i^\alpha}\right\rangle\,. 
    \]
    These coordinates are called \textit{Darboux coordinates} of the polarised $k$-contact manifold $(M,\mathbf{\eta},\mathcal{V})$.
\end{theorem}

Theorem \ref{Th::PolkCon} allows us to consider the manifold introduced in Example \ref{ex:canonical-k-contact-structure} as the canonical model of polarised    $k$-contact manifolds. Moreover, every polarised $k$-contact manifold that is the contactification of a polarised $k$-symplectic manifold has Darboux coordinates.

\subsection{$k$--Contact Hamiltonian systems}
\label{sub:k-contact-Hamiltonian-systems}

Let us present the basics of the Hamiltonian $k$-contact formulation of field theories (for more details see \cite{GGMRR_20}).


\begin{definition}
    Consider a $k$-contact Hamiltonian system $(M,\mathbf{\eta},h)$, that is a $k$-contact manifold $(M,\mathbf{\eta})$ and a Hamiltonian function $H\in\mathcal{C}^\infty(M)$. The \textit{$k$-contact Hamilton--De Donder--Weyl equations} for a $k$-vector field $\mathbf{X} = (X_1,\ldots, X_k)\in\mathfrak{X}^k(M)$ are
    \begin{equation}\label{eq:k-contact-HDW-fields}
            \iota_{X_\alpha}d\eta^\alpha = d h - (R_\alpha h)\eta^\alpha\,,\qquad
            \iota_{X_\alpha}\eta^\alpha = -h\,.
    \end{equation} 

    A $k$-vector field $\mathbf{X}$ that satisfies equations (\ref{eq:k-contact-HDW-fields}) is called a \textit{$k$-contact Hamiltonian $k$-vector field}. A $k$-contact Hamiltonian system gives rise to a family of $k$-contact Hamiltonian vector fields. 
\end{definition}

In Darboux coordinates, if  $
    X_\alpha = (X_\alpha)^i\parder{}{q^i} + (X_\alpha)^\beta_i\parder{}{p_i^\beta} + (X_\alpha)^\beta\parder{}{z^\beta}\,,\;\alpha=1,\ldots,k,$ then, equations (\ref{eq:k-contact-HDW-fields}) are equivalent to the following relations
\begin{equation}\label{eq:k-contact-HDW-fields-Darboux-coordinates}
        (X_\alpha)^i = \parder{h}{p_i^\alpha}\,,\qquad
        (X_\alpha)^\alpha_i = -\left( \parder{h}{q^i} + p_i^\alpha\parder{h}{z^\alpha} \right)\,,\qquad
        (X_\alpha)^\alpha = p_i^\alpha\parder{h}{p_i^\alpha} - h\,.
\end{equation}

\section{A relation between Hamiltonian $k$-symplectic systems and Hamiltonian $k$-contact systems}

Every polarised exact $k$-symplectic manifold allows one to define a polarised $k$-contact manifold. It is natural to explore whether this relation extends to the corresponding Hamiltonian systems. That is, if we consider a $k$-contact Hamiltonian system where the $k$-contact manifold is derived from the contactification of a polarised exact $k$-symplectic manifold, can the solutions of the $k$-symplectic Hamiltonian system be recovered from the solutions of the $k$-contact Hamiltonian system?

\begin{proposition}\label{prop:relation_formalism}
    Let $(P,\mathbf{\omega},h)$ be a $k$-symplectic Hamiltonian system defined on a   polarised exact $k$-symplectic manifold $(P,\mathbf{\omega}=d\mathbf{\theta},\mathcal{V})$. 
Let $(M=P\times\mathbb{R}^k,\mathbf{\eta}_M,\mathcal{V}_M)$ be the polarized $k$-contact manifold defined by the contactification of $(P,\omega)$. Then $(M,\mathbf{\eta}_M,h_M=pr_1^*h)$ is a $k$-contact Hamiltonian system, where $pr_1\colon M\to P$ is the natural projection.

Moreover, if $\mathbf{X}_M\in\mathfrak{X}^k(M)$ is a $pr_1$-projectable solution of (\ref{eq:k-contact-HDW-fields}), then its projection $\mathbf{X}_P\in \mathfrak{X}^k(P)$ is a solution of  (\ref{eq::k-sym-HDW-FE}).
\end{proposition}
\begin{proof}
    Let $(P,\mathbf{\omega}=d\mathbf{\theta},h)$ be a polarized exact $k$-symplectic Hamiltonian system.  We consider $(M,\eta_M,\mathcal{V}_M)$ the contactification of $(P,\omega=d\theta,\mathcal{V})$ (see Example \ref{ex:contactification-k-symplectic-manifold}). 
    Given the Hamiltonian function $h\colon P\to \mathbb{R}$ we define $h_M\colon =(pr_1)^*h$, that is,  $h_M(m)=h(pr_1(m))=h(p)$, where $m=(p,x)\in M=P\times\mathbb{R}^k$. Thus we obtain a $k$-contact Hamiltonian system $(M,\mathbb{\eta}_M,h_M)$

   Let $\mathbf{X}_M=(X_1,\ldots, X_k)$ be a $pr_1$-projectable $k$-vector field solution of (\ref{eq:k-contact-HDW-fields})  and $\mathbf{X}_P=(Tpr_1(X_1),\ldots, Tpr_1(X_k))$  its projection. We have 
    \[
    \iota_{X_\alpha}d\eta^\alpha = dh_M -(R_\alpha)(h_M)\eta^\alpha = dh_M=dpr_1^*(h) = pr_1^*(dh)\,.
    \]
    On the other hand,
    \[
    \iota_{X_\alpha}d\eta^\alpha =\iota_{X_\alpha}d(dz^\alpha+\theta_M^\alpha)=\iota_{X_\alpha}d\theta_M^\alpha=\iota_{X_\alpha}d(pr_1)^*\theta^\alpha = pr_1^*(\iota_{Tpr_1(X_\alpha)}d\theta^\alpha)
    \]
    Therefore, $\iota_{Tpr_1(X_\alpha)}d\theta^\alpha=dh$, that is, $Tpr_1(\mathbf{X}_M)=(Tpr_1(X_1),\ldots, Tpr_1(X_k))$ is a solution of  (\ref{eq::k-sym-HDW-FE}).
\end{proof}

\begin{example}[The  wave equation] A vibrating string can be described using a $k$-contact Hamiltonian formalism \cite{LRVZ25,GGMRR_20}. In the particular case of the undamped vibrating string problem, this can be described also using the $k$-symplectic formalism.  
    
    Consider the coordinates $\{t,x\}$ for $\mathbb{R}^2$ and set $Q=\mathbb{R}$. The phase space becomes $\bigoplus^2\mathrm{T}^\ast \mathbb{R}$ and accepts the coordinates $(u,p^t,p^x)$. Denote by $u$ the separation of a point in the string from its equilibrium point, while $p^t$ and $p^x$ will denote the momenta of $u$ with respect to the two independent variables. We endow $\bigoplus^2\mathrm{T}^\ast \mathbb{R}$ with its natural two-symplectic form $\omega=d\theta$, see Example \ref{ex:canonical-model-k-symplectic}. A Hamiltonian function for the vibrating string can be chosen to be $h\in\mathcal{C}^\infty ( \bigoplus^2\mathrm{T}^\ast \mathbb{R})$ of the form
\begin{equation}\label{eq:hamiltonian-vibrating-string}
    h(u,p^t,p^x) = \frac{1}{2\rho}(p^t)^2 - \frac{1}{2\tau}(p^x)^2,
\end{equation}
where $\rho$ is the linear mass density of the string and $\tau$ is the tension of the string. We assume that $\rho$ and $\tau$ are constant. 

We now consider $(M=\bigoplus^2\mathrm{T}^\ast \mathbb{R}\times\mathbb{R}^2,\eta_M)$ the contactification of the previous $2$-symplectic structure (see example \ref{ex:contactification-k-symplectic-manifold}) and the Hamiltonian function $h_M=(pr_1)^*h$. Let us observe that this Hamiltonian function is the Hamiltonian function of the damped vibrating string problem described in \cite{LRVZ25} in the particular case $k=0$, where $k$ denoted the damped constant. We consider an integrable $pr_1$-projectable $2$-vector field solution of (\ref{eq:k-contact-HDW-fields})(see local expression in \cite{LRVZ25}, then from Proposition \ref{prop:relation_formalism} we obtain that a solution of the equation (\ref{eq::k-sym-HDW-FE}) is the $2$-symplectic Hamiltonian $2$-vector field given by 
\[
Tpr_1(X^1)=\frac{p^t}{\rho}\parder{}{u}+A^1_t\parder{}{p^t}+A^1_x\parder{}{p^x},\quad 
Tpr_1(X^2)=-\frac{p^x}{\tau}\parder{}{u}+A^2_t\parder{}{p^t}-A^1_t\parder{}{p^x},\]
where $A^1_t,A^1_x,B^1_t,B^1_x$ are arbitrary functions on $M$ such that $A^1_t+A^2_x=0$. Thus, the Hamilton--De Donder--Weyl field equations are
\begin{equation}
  		\parder{u}{t} = \frac{1}{\rho}p^t\,,\quad
 		\parder{u}{x} = -\frac{1}{\tau}p^x\,,\quad
 		\parder{p^t}{t} + \parder{p^x}{x} = 0\,,\\
 \end{equation}
Hence, by substituting the first and second equations in the third, one obtains
$
    \frac{\partial^2 u}{\partial t^2}-c^2 \frac{\partial^2 u}{\partial x^2}  = 0 \,,
$
where $c^2 = \frac{\tau}{\rho}$, that is, the undamped wave equation.
\end{example}

\section{Conclusions and outlook}
In this paper a first relation between $k$-symplectic Hamiltonian systems and $k$-contact Hamiltonian systems has been studied.  The future idea is to continue this study by considering new situations in which both types of systems can be related. In order to analyse a converse of the result of Proposition \ref{prop:relation_formalism}, it is important to note that not every $k$-contact manifold is the contactification of an exact $k$-symplectic manifold so that an equivalence between both results does not make sense. However, it is possible to raise questions in order to provide partial converse of this result, which are proposed as future work. In particular, if the $k$-contact manifold is the contactification of an exact $k$-symplectic manifold, we consider the question: is it possible to recover all the $k$-contact Hamiltonian $k$-vector fields solutions of equation (\ref{eq:k-contact-HDW-fields}) from the all set of all $k$-symplectic Hamiltonian $k$-vector fields solutions of \ref{eq::k-sym-HDW-FE}. We think that the answer will be negative since it is possible that there are solutions of (\ref{eq:k-contact-HDW-fields}) which are not $pr_1$-projectable. On the other hand, the Theorem 4.10 in \cite{LRVZ25} shows how a $k$-contact manifold can be extended to a $k$-symplectic manifold, and vice versa. As a future work we consider the study of the relation between $k$-contact Hamiltonian $k$-vector field and $k$-symplectic Hamiltonian $k$-vector field in this context.

\begin{credits}
\subsubsection{\ackname} 
S. Vilariño acknowledges partial finacial support from the Spanish Ministry of Science and Innovation, grant PID2021-125515NB-C22 and Aragon Government, project 2023 E48-23R.

\subsubsection{\discintname}
The authors have no competing interests to declare that are
relevant to the content of this article.
\end{credits}

\end{document}